# Theory and Practice of Data Citation

Gianmaria Silvello

Department of Information Engineering, University of Padua, Via Gradenigo 6/b, Padua, Italy

silvello@dei.unipd.it

tel. +39 049 827 7500

## Abstract

Citations are the cornerstone of knowledge propagation and the primary means of assessing the quality of research, as well as directing investments in science. Science is increasingly becoming "data-intensive", where large volumes of data are collected and analyzed to discover complex patterns through simulations and experiments, and most scientific reference works have been replaced by online curated datasets. Yet, given a dataset, there is no quantitative, consistent and established way of knowing how it has been used over time, who contributed to its curation, what results have been yielded or what value it has.

The development of a theory and practice of data citation is fundamental for considering data as first-class research objects with the same relevance and centrality of traditional scientific products. Many works in recent years have discussed data citation from different viewpoints: illustrating why data citation is needed, defining the principles and outlining recommendations for data citation systems, and providing computational methods for addressing specific issues of data citation.

The current panorama is many-faceted and an overall view that brings together diverse aspects of this topic is still missing. Therefore, this paper aims to describe the lay of the land for data citation, both from the theoretical (the why and what) and the practical (the how) angle.

## Introduction

Citations are the cornerstone of knowledge propagation in science, the principal means of assessing the quality of research and directing investments in science as well as one of the pillars of the scholarship architecture.

Currently we are witnessing a crucial change in the way research is conducted and science progresses (Bechhofer *et al.*, 2013). We are rapidly transitioning towards *the fourth paradigm of science* (i.e. data-intensive scientific discovery), where data are as vital to scientific progress as traditional publications are. Data and scholarship are increasingly interwoven and new concepts such as "data scholarship" (Borgman, 2015) and "data-intensive research" (Hey & Trefethen, 2005) are becoming popular and central to the academic and scientific world.

Experimental and observational data and scientific models are born digital, and this move to digital content has to be taken into account by scholarly publications, credit attribution processes and scientific and economic impact measurement. Vast amounts of scientific data – molecular, geospatial, astronomical, pharmacological and more – are now collected and made available in structured, evolving and often distributed databases (Buneman, Davidson & Frew, 2016); relevant examples of widely used and accessible scientific databases are the pharmacological IUPHAR/BPS[1] database, the Eagle-i[2] biomedical dataset, the DrugBank[3] bioinformatics and cheminformatics database, the Reactome[4] pathways database and the VADMC[5] federated database of atomic and molecular data. Moreover, the scientific community is taking action to promote an *open research culture* (Nosek *et al.*, 2016) and effective means to share, discover and access scientific data that too often still remain "hidden" at their origin or are shared in sub-optimal ways (Ohno-machado *et al.*, 2015). Indeed, as a matter of good scientific practice, scientific databases are recommended to be *FAIR* (Findable, Accessible, Interoperable, and Reusable) (Wilkinson *et al.*, 2016) and accessible from the articles referring to or using it (Cousijn *et al.,* 2017).

---

[1] http://www.guidetopharmacology.org/
[2] https://www.eagle-i.net/
[3] https://www.drugbank.ca/
[4] http://www.reactome.org/

Since scientific works and publications increasingly rely on curated databases, traditional references are starting to be placed alongside references to data; hence, scientific publishers (e.g. Elsevier, PLoS, Springer, Nature) have been defining data policies and author guidelines to include data citations in the reference lists (Cousijn *et al.,* 2017; Walton, 2010). At the institutional level, within the Horizon 2020 program, the European Commission has introduced the Open Research Data Pilot (ODP), which aims to improve and maximize the access to and re-use of research data and to increase the credit given to data creators; in the same vein, also the National Science Foundation (NSF) and the National Institutes of Health (NIH) in the United States and the Economic and Social Research Council in the United Kingdom ask researchers applying for grants to provide data management plans describing how the data they use and produce will be shared and referenced (MacKenzie, 2012; Spengler, 2012). There is also evidence that: (i) scientists respond to incentives and that increasing citation would drive software development and data sharing (Niemeyer, Smith & Katz, 2016); (ii) research activity increases when outputs are formally counted (McNaught, 2015); and (iii) direct citations to datasets stimulate more data curation and data sharing than indirect citations through publications (Belter, 2014). Moreover, fundamental aspects of scientific research, such as reproducibility of experiments, the availability and discovery of scientific data and the connection between scientific results with the data providing evidence, have been found to be closely connected with data citation (Honor, Haselgrove, Frazier & Kennedy, 2016).

As a consequence, there is a strong demand (CODATA, 2013; FORCE11, 2014; RDA, 2015) to give databases the same scholarly status of traditional references and to define a shared methodology to cite data.

Nevertheless, none of the largest citation-based systems – i.e. Elsevier Scopus, Thompson Reuters Web of Science, Microsoft Academia and Google Scholar – consistently take into account scientific datasets as targeted objects for use in academic work. The Thompson Reuters (now Clarivate Analytics) Data Citation Index (DCI) (Force, Robinson, Matthews, Auld & Boletta, 2016) is one notable exception that aggregates information about the usage of data. The DCI is an infrastructure that keeps track of data usage in the scientific domain, which indexes a number of scientific data sets, allows for discovering the data sets selected and validated by Thompson Reuters and provides the technical means to connect entire data sets or repositories to scientific papers. On the other hand, DCI is still in its infancy, "has yet to be proven cost-effective or efficient" (Mayernik, Hart, Mauli & Weber, 2016) and presents skewed citation patterns (Robinson-Garcia, Jiménez-Contreras & Torres-Salinas, 2016). Therefore, given a dataset, we still not have any quantitative, consistent and established way of knowing how it has been used over time, what analyses have been conducted on it, what results have been yielded, who contributed to its curation and which data subsets have been relevant or impactful for scientific and economic development (Honor *et al.*, 2016).

This is mainly due to the lack of a "deep and persistent mechanism for citing data" (Ingwersen & Chavan, 2011). Indeed, citation rules and practices have stratified over centuries and are well established for text but less so for data (Borgman, 2015). Data citations and traditional paper-based citations have a common intent, but present some epistemological differences that have not yet been fully analyzed and comprehended. Among other things, these differences impact the very use of data citations in bibliometrics (and more generally in infometrics) as pointed out by

---

Borgman (2016): "[the] leap from citing publications to citing data is a vast one [and the transfer of] bibliographic citation principles to data must be done carefully and selectively". Indeed, bibliometrics is based on publications in the sense of scholarly communications and data do not necessarily qualify as such (Mayernik, Callaghan, Leigh, Tedds & Worley, 2015). Bibliometrics relies on the distinction between reference and citation where a "reference is the acknowledgement that one document gives to another; a citation is the acknowledgement that one document receives from another" (Narin, 1976). Whereas, in data citation, this distinction is often blurred and reference and citation are treated as identical signs. Moreover, another distinctive feature of bibliometrics are reciprocal citations, which are hardly possible with data citations since data may receive citations but cannot cite something else.

Nevertheless, data citations are often referred to as "formal ways to ground the research findings in a manuscript, upon their supporting evidence, when that evidence consists of externally archived datasets" (Cousijn *et al.*, 2017) and it is argued that citations to data should be placed in the reference list section as published works are (Altman, 2012; Corti, Van den Eynden, Bishop & Woollard, 2014; Park & Wolfram, 2017). Hence, a data citation could be indicated with the triple {R, C, D}, where $R$ is the reference paper, $C$ is the citation and $D$ is the data set – i.e. the object of the citation. In this context, the major practical issues for citing data, regard the definition and the creation of the text snippet to be included in the reference list of $R$ and the identification and retrieval of $D$. In addition, we must consider that data are more complex and varied than documents (Castelli, Manghi & Thanos, 2013; Wynholds, Wallis, Borgman, Sands & Traweek, 2012) and they introduce new challenges with respect to traditional publications. Indeed, text publications have a fixed form, do not change over time, are interpretable as independent units, share a common format and representation model and are composed of predetermined, albeit domain-dependent, sets of citable units (e.g. authors, title, pages). By contrast, scientific datasets are structured according to diverse data models – e.g. relational, hierarchical (XML), graph-based (RDF) – and accessed with specific query languages – e.g. SQL, XPath/XQuery, SPARQL. Data citable units range from a single datum to data subsets or aggregations specified on the fly with great variability, and deciding *a priori* what can be cited and what cannot, is not (always) feasible. Data evolve over time, but data citations must be *diachronic*: they must be consistent over time and always lead back to the original cited data. A data reference (or citation snippet) is required to allow the data to be understood and correctly interpreted and it must be composed of the essential information for identifying the cited data as well as contextual information. Such contextual information must be extracted from the given dataset and/or from external sources automatically, because we cannot assume one knows how to access and select additional relevant data and to structure them appropriately (Hourclé, 2012). The role of contextual information is of critical importance if we consider the potential, albeit undesirable, situation where data references may survive to the data themselves.

Data citation is a complex problem having both theoretical/conceptual and practical/technical implications that have been investigated both by information science and computer science. The former discipline has been mainly concerned with the motivations – the "why" of data citation – as well as with the principles of this field and the main features of citation systems expressed in terms of recommendations – the "what" of data citation. The latter discipline has been mainly concerned with the computational problems arising from data citation and the design of systems for automating the citation of data – the "how" of data citation.

The field of data citation can be likened to a broken mirror where each single piece of glass reflects a different facet of the problem. Therefore, it is difficult to get the complete picture of the matter, which is nevertheless fundamental for providing a shared theory of data citation as well as a sound data citation system. This paper aims to describe the lay of the land for data citation both from the theoretical (the why and what) and the practical (the how) angle. Moreover, we provide an overview of the most pressing open problems in data citation as well as some related research directions for information and computer science.

*A note on literature search*

Literature search was conducted as a four-stage process. The first stage was a journal database search based on key term search in Google Scholar and Scopus. We searched for the exact key phrase "data citation" and we discarded most of the papers regarding citation data analysis which are out of scope for this review. We analyzed articles that cover data citation as a need/service/tool to be employed in a specific domain (e.g. biology or pharmacology) and articles discussing the principles and the technical aspects of data citation. We did not consider articles about data publishing, sharing and re-use if they did not explicitly mention data citation.

In the second stage, we analyzed articles published in discipline-specific venues (both journals and conferences), in particular we considered information science, digital libraries and computer science with a specific focus on information management and access systems. In the third stage, we conducted a general search for the "data citation" key-phrase by employing general-purpose Web search engines. This allowed us to look for relevant blog entries, white papers and publisher web pages. As a final stage, we scanned (recursively) the reference list of the papers selected in the previous stages for additional relevant items.

## Why data citation: Motivations

The main motivation advocating the importance of data citation is that "*data in research are as valuable as papers and monographs*" (Ball & Duke, 2012); moreover, data citation is becoming central in science because of the great production of data, new research tools for managing, accessing, analyzing and sharing data and the shift of research policy that has been happening in recent years (Borgman, 2012a). The centrality of data citation is widely recognized across many scientific fields ranging from earth sciences, physics and biology to information studies and social sciences. The need for a standard citation mechanisms to give credit to data collectors and to locate and examine the data used in investigations is a long-standing requirement also in the computer science area, especially for database management (French, Jones & Pfaltz, 1990).

Data citation has many facets and it is needed to satisfy several requirements depending at times on the specific discipline in question. Despite this heterogeneity, we have identified six main motivations for data citations that are shared in many different scientific fields:

i.   *Data attribution*: Scientific data are collected and curated with a great deal of work and in many cases this is done directly by qualified scientists and researchers. Giving credit to the people involved in the creation and curation of data is important for scientific recognition (Tenopir *et al.*, 2011). Data attribution opens up the need to identify the author or the person responsible for data with variable granularity; indeed, in many fields it is not enough to give credit to the person who manages the dataset as a whole, as there may be different groups of authors for different subsets of data within the same dataset. Parsons (2012) and Borgman (2015) relate the concept of attribution with that of accountability for the data since if we can identify who is responsible for the data, we know who gets the merit for the data as well as who can be held accountable for them.

ii.  *Data connection*: Data citation methods are required to connect scientific papers with the underlying data on which they are based. Castelli, Manchi & Thanos (2013) indicate data citation as the "main mechanism enabling the alignment and integration between data and publications in the scientific communication process". This is also the idea behind executable papers or enhanced publications (Vernooy-Gerritsen, 2009) which link data with textual publications allowing the data to be consulted or downloaded while reading a scientific paper in a digital form as well. Several studies (Aalbersberg, Heeman, Koers & Zudilova-Seinstra, 2012; Attwood *et al.*, 2010; Bardi & Manghi, 2015; Brammer, Crosby, Matthews & Williams, 2011; Jankowski, Scharnhorst, Tatum & Tatum, 2012) have been dedicated to the definition and realization of enhanced publications and they all to some extent require a methodology for citing data (and software/code) (Niemeyer, Smith & Katz, 2016). There is also the emerging idea of using data citation in conjunction with the Linked Data paradigm to create a "claim network evidence" spanning different documents (de Waard, 2012; Silvello, 2015). The idea is to backup claims in scientific papers with the data providing evidence for their validity.

iii. *Data discovery*: Being able to cite data means being able to identify, reach, access and retrieve a dataset or a subset of a dataset, which are the fundamental operations required to make data discoverable. Furthermore, data citations may act as entry points to hidden data sources which are not indexed by search engines and thus are virtually unreachable to all those scientists who do not have the knowledge and the means to reach and interrogate these data sources.

iv. *Data sharing*: The possibility of citing data and thus of giving credit to data scientists and institutions is considered a decisive factor for augmenting the willingness of scientists to share data. Indeed, since the technological factor is no longer an impediment for data sharing, the main barrier still standing is the fear scientists have of losing a competitive advantage while "receiving no credit and losing funding or publishing opportunities" (Mooney & Newton, 2012). Data sharing is a condition for enabling data discovery and data re-use as well as one of the main factors for the success of the Linked Data paradigm in the scientific context. To this end, Bechhofer *et al.* (2013) highlighted that Linked Data is a compelling approach for sharing and disseminating scientific data, but in order to be effective for data re-use and discovery it has to be connected to the research methodology and respect the rights and the reputation of the researchers.

v. *Data impact*: Citing data allows us to define new usage metrics for determining the impact of data; data impact can be interpreted as a way to measure or discover which results have been obtained using the data, how many times they have been used, where they have been used and so on. Note that in the literature, data impact is, frequently, intended as the rate of data usage and not as a quality measure. This aspect is particularly important for public and private funding agencies that often invest in the creation of scientific dataset as well as research infrastructures; data citation is required to estimate the use of data going beyond alternative metrics such as altmetrics (Mayernik, 2016). de Waard (2016) pointed out that "making data citable" is one of the ten main characteristics for unlocking the potential of research data; in particular, being able to cite data is important for increasing research exposure. There are also voices calling for considering data citations for bibliometrics with the purpose of evaluating research performances of institutions and individuals (Costello, 2009; Parr, 2007). Moreover, funding agencies underline the need to cite data for measuring the impact of researchers working with data (Ahalt et al., 2015) in order to "help people in career advancements and make their contribution clear" (Spengler, 2012). On the other hand, the use of citation data (e.g. citation count and citation indexes and databases) for evaluating scholarship and funding distribution also raises ethical and policy issues (Borgman, 2016; Garfield & Welljams-Dorof, 1992; Furner, 2014) when applied both to traditional and data citations. In particular, for data citations, given the current absence of "norms of citing behavior", we have to be even more careful by considering that they cannot be straightforwardly analogized to paper citations (Stuart, 2017).

vi. *Reproducibility*: Data citation has a profound impact on the reproducibility of science (Baggerly, 2010), a hot topic in many disciplines such as astronomy (Kurtz, 2012), biology (Bloom, Ganly & Winker, 2014), physics, computer science (Ferro, 2017; Freire, Fuhr & Rauber, 2016) and more. Lately, several authoritative journals have been requesting the sharing of data and the provision of validation methodologies for experiments (e.g. Nature Scientific Data and Nature Physics); these publications and the publishing industry in general see data citation as the means for providing new, reliable and usable means for sharing and referring to scientific data.

| | Data Attribution | Data Connection | Data Discovery | Data Sharing | Data Impact | Reproducibility | Application Domain |
|---|---|---|---|---|---|---|---|
| (Altman & King, 2007) | ✔ | | | | | ✔ | Information Studies |
| (Altman, 2012) | ✔ | | ✔ | | | | Information Studies |
| (Arend et al., 2016) | | | | ✔ | ✔ | | Biology |
| (Ball & Duke, 2015) | ✔ | ✔ | ✔ | ✔ | | ✔ | Information Studies |
| (Bardi & Manghi, 2014) | ✔ | ✔ | ✔ | ✔ | | ✔ | Data Publishing |
| (Belter, 2014) | ✔ | | | | ✔ | ✔ | Oceanography |
| (Borgman, 2012) | ✔ | ✔ | ✔ | | | | Information Studies |
| (Borgman, 2015) | ✔ | ✔ | ✔ | ✔ | | | Information Studies |
| (Bravo et al., 2015) | ✔ | | ✔ | ✔ | ✔ | ✔ | Biodiversity |
| (Callaghan et al., 2012) | ✔ | | | ✔ | | | Environmental Science |
| (Candela et al., 2015) | | | ✔ | ✔ | | | Data Publishing |
| (Chavan, 2012) | | | ✔ | ✔ | | | Biodiversity |
| (Cook et al., 2016) | ✔ | ✔ | ✔ | ✔ | | | Environmental Science |
| (Costello et al., 2013) | | | | ✔ | ✔ | | Biodiversity |
| (de Waard, Cousijn & Aalbersberg, 2015) | ✔ | | | ✔ | ✔ | | Data Publishing |
| (Dodd, 1979) | | | ✔ | ✔ | ✔ | | Social Sciences |
| (Duerr et al., 2011) | ✔ | | | | | ✔ | Earth Science |
| (Edmunds et al., 2012) | ✔ | | | ✔ | | ✔ | Biology |
| (Fecher, Friesike &Hebing, 2015) | ✔ | | | ✔ | ✔ | | Data Management |
| (Herterich et al., 2016) | ✔ | | | ✔ | | | Physics |
| (Honor et al., 2016) | ✔ | | ✔ | ✔ | ✔ | ✔ | Neuroimaging |
| (Huang, 2015) | ✔ | | | ✔ | | | Crystallography |
| (Ingwersen & Chavan, 2011) | | | | | ✔ | | Bioinformatics |
| (Kafkas, Kim, Pi & McEntyre, 2015) | | ✔ | | | | | Biomedicine |
| (King, 1995) | | | | | | ✔ | Information Studies |
| (Klump, Huber & Diepenbroek, 2016) | | | | | | ✔ | Earth Science |
| (Koers, 2015) | ✔ | | | | | ✔ | Clinical Epidemiology |
| (Mayernik, 2012) | ✔ | ✔ | | | ✔ | ✔ | Information Studies |
| (Mayernik, 2016) | | | | | ✔ | ✔ | Information Studies |
| (Mooney & Newton, 2012) | ✔ | | ✔ | ✔ | | | Data Publishing |
| (Nature Physics, 2016) | ✔ | | | | | | Physics |
| (Parson, 2012) | ✔ | | | | ✔ | ✔ | Earth Science |
| (Parsons & Fox, 2013) | ✔ | | | | | ✔ | Data Publishing |
| (Peters, 2016) | | | | | ✔ | | Information Studies |
| (Pröll & Rauber, 2013) | | | | | | ✔ | Computer Science |
| (Sieber & Trumbo, 1995) | ✔ | | | ✔ | ✔ | ✔ | Information Studies |
| (Silvello & Ferro, 2016) | ✔ | ✔ | | | | ✔ | Digital Libraries |
| (Silvello, 2015) | ✔ | ✔ | | | | ✔ | Information Retrieval |
| (Silvello, 2017) | ✔ | ✔ | | | | ✔ | Computer Science |
| (Simons, Visser & Searle, 2013) | ✔ | | | ✔ | | ✔ | Digital Libraries |
| (Starr et al., 2015) | | ✔ | | | ✔ | | Computer Science |
| (Thorisson, 2009) | ✔ | ✔ | | | | | Biotechnology |
| (White, 1982) | ✔ | | ✔ | | ✔ | ✔ | Social Sciences |
| (Wormack, 2015) | | | ✔ | ✔ | | | Information Studies |
| (Zwolf et al., 2016) | ✔ | | | ✔ | ✔ | ✔ | Spectroscopy |

Table I Summary of the main motivations for data citation in different application domains

In Table I we present a summary of the main motivations for data citation as presented by several papers discussing this topic from different perspectives and application domains. We can see that data attribution and reproducibility are the core topics motivating data citation, followed by data sharing. Data connection has been gaining traction in recent years as it is closely related to reproducibility. Data discovery and data impact are growing and becoming central motivations for data citation, especially within the information studies, data publication and data curation fields.

These motivations explain why data citation is needed for scholarly communications and they constitute a *prior analysis* of why data should/must be cited. From these motivations, though, we cannot infer the reasons why authors decide to cite data and their data citing behavior. To this end, we should reconsider the meaning of data citations, as it may differ from textual citations (Mayernik, 2012) and "we should ask whether the motivations for citing data are the same as for citing research papers" (Robinson-Garcia, Jiménez-Contreras & Torres-Salinas, 2016). In the context of traditional publications, there are different theories of citation (Leydesdorff, 1998) analyzing why we cite (Cozzens, 1981) and when we cite (Garfield, 1996). In particular, it has been pointed out that citations can be seen from a normative perspective as ways to acknowledge intellectual debt (Kaplan, 1965). Alternatively, citations are seen as a means for persuasion, since scientific papers have the role of convincing people of the validity of claims and providing support for them (Gilbert, 1977); or as symbols relating the author's own private interpretation of the reference with the cited paper itself (Small, 1978). Garfield (1996) listed fifteen major reasons for citation and Cronin (1984) reported several alternative citation categories and typologies defined over the years, to conclude that there is reason to believe that "there are no absolute and exclusive categories which fully describe the relationship of a citing publication to the cited publication" (Lipetz, 1965). Nonetheless, textual analysis of the citing paper could suggest a plausible explanation about why an author cites as s/he does (Cronin, 1981); moreover, given that the citation process is "subjective and inhospitable to standardization" (Cronin, 1981), we have to rely on "ostensible reasons for citation or reasons which can be adduced from the context of the citing work" (Frost, 1979).

For data citation, we can find just a few works specifically considering the context of the citing work (Kafkas, Kim & McEntyre, 2013; Piwowar & Vision, 2013) in order to understand why data are cited. This is also due to the fact that data citations "are useful for the discovery of links, but less so for understanding why the link was made" (Mayernik Philips & Nienhouse, 2015). Mayernik *et al.* (2014) are among the few hypothesizing that data may be cited as "forms of reward to data providers", as a "quid pro quo for the data provider support" or as a "persuasive statement illustrating how data of high visibility or sufficient quality underlie a scientific result". Up to now, we can only find isolated and sporadic efforts to understand why data are cited. However, a shared effort for defining a theory of data citing is very much needed, especially before we could "safely" employ bibliometrics for data citations given that "with limited understanding of the norms of data publishing and data citation, too early an emphasis on metrics may do damage to the nascent data ecosystem" (Stuart, 2017).

To this end, we may recall that the development of a citation theory (and the study of citations as a serious academic subject) emerged as a consequence of the successful development of commercial citation indexes (Cronin, 1984). Hence, for data citation, the Thomson-Reuters DCI (and other indexes that may emerge in the next years) can play a role towards the study of a theory of data citation. In particular, the DCI has been used to investigate why data are cited in Genetics and Heredity (Park & Wolfram, 2017) and in the Humanities (Robinson-Garcia, Jiménez-Contreras & Torres-Salinas, 2016). A similar study was carried out, long before the DCI was established, by White (1982) who conducted an analysis of data citations in the context of social sciences. All these studies concluded that the acknowledgement of data usage is the prior motivations for citing a dataset. Other studies (Fear, 2013; Kafkas, Kim & McEntyre, 2013; Weber, Mayernik & Worley, 2014) corroborated this result by suggesting that the number of citations to a dataset correlates with its usage rate. Thus, from these initial and circumscribed analyses, it emerges that a strong motivation for an author to cite a dataset is to acknowledge the fact that she used it in her research. Having said that, it is complex to define what is meant by "usage" and data citation requires "a more nuanced understanding of data "use" to be effective" (Wynholds, Wallis, Borgman, Sands & Traweek, 2012).

## What data citation: Principles and system requirements

The principles of data citation have been extensively described in (CODATA, 2013) and then aggregated, summarized, revised and endorsed by the Joint Declaration of Data Citation Principles (JDDCP) (FORCE11, 2014). These principles can be classified into two main groups: the former states the role of data citation in scholarly and research activities and the latter defines the main guidelines a data citation methodology and system should respect.

In the first group, we find *importance*, *credit and attribution*, *evidence*, *verifiability* and *interoperability*. Importance coincides with the first motivation for data citation we presented above, which is that data should be considered as a first-class citizen in the science panorama and thus be treated as any other scholarly record and scientific paper. This principle is foundational for data citation, because it opens up to the use of data for assessing the work of individuals and institutions, for defining new impact measures and for directing investments in research. This calls for a change of policy in research that invests in public institutions as well as in publishing houses and individual researchers. Furthermore, the principle of importance of data opens up a plethora of theoretical issues that have been only partially discussed in the literature and call for new research; Borgman (2016) points out several theoretical problems of data citation mainly due to the lack of agreement of what constitutes data, amongst which the fact that data are different from traditional publications and require a new set of theoretical premises for bibliometrics considering how citation practices differ between genres of publications and data. Moreover, there is not a shared vision about which data could or should by cited also considering that different application domains call for different data citation practices accounting for references to primary data sources, raw data (e.g., sensor or streaming data), post-processing data, or formatted data as tables and plots.

The JDDCP principles of credit and attribution, evidence and verifiability correspond to the some of the motivations for data citation identified above; indeed, citation to data should give credit to people and institutions involved in their creation and curation and scientific claims should be related by the specific data supporting them. Within traditional publications, the accepted protocol for credit and attribution is for a publication to include "in-text citations and a reference section the lists those cited publications" (Honor *et al.*, 2016). Also in the context of data citation, every citation should come with a citation text – i.e. citation snippet or reference text – which describes the object being cited and should contain enough information to give credit to data creators and curators, to understand the motive of the citation and the meaning of the cited object, to locate the object referred to as evidence, and to verify that the located object is semantically identical to the cited one (Altman & King, 2007; Altman, 2012). The citation snippet should be human- and machine-readable; human-readable snippets are required to give immediate sense of a citation even to non-experts without requiring access to the data being referred to by the citation (Kafkas, Kim, Pi & McEntyre, 2015; Silvello & Ferro, 2016; Van de Sompel, 2012). Machine-readable snippets are required for automatic processing of the citation information and they can contain more exhaustive information than human-readable ones; for instance, we may have hundreds of contributors for a given dataset that can be exhaustively reported only in a machine-readable snippet. To this end, a fundamental requirement for citation methods is to automatically create citation snippets in order to ensure consistency of references (Thorisson, 2009). Indeed, recent studies about data citation practices showed that data citation snippets are often underspecified with vital information missing (Mathiak & Boland, 2015) and may present several errors compromising the identification and retrieval of the data being cited (Henderson & Kotz, 2015). *Automatic methods for building citation snippets* are also required because humans cannot remember or know what the necessary information is, when a snippet is complete, what the citation format is required in a given context and where to gather the relevant information, especially in a big data context where manual exploration of data is unfeasible (Buneman *et al.*, 2014; Crosas *et al.*, 2015; Jagadish, 2015; Minister, 2012; Peters, 2016). The problem of automatic citation snippet creation has been clearly defined by (Buneman *et al.*, 2016) from a database management system perspective: "*Given a database D and a query Q, generate an appropriate citation*"; in this context, the term database and query are used in a very general way to refer to any mechanism used to extract the data from a generic source. Borgman (2012; 2016), Buneman *et al.* (2000) and Groth (2012) also outlined the relationship between provenance and data citation. Provenance information should be included in a citation snippet in order to cite the correct version, manipulation and transformation of data. This aspect is of particular relevance to cite meta-analyses conducted on raw or curated data in order to cite and keep track of the original dataset as well as of the transformations and manipulations of the data. The consistency and completeness of citation snippets within a discipline is a concern also from the metadata viewpoint. Indeed, Uhlir (2012) underlines the importance of reference consistency since, for instance, (Bolter, 2014) reported cases where there are several hundred variants of citations to the same dataset (i.e. the World Ocean Atlas and World Ocean Database). Borgman (2012b) and NSF (2015) stressed the importance of developing shared and extensible metadata formats for data citation. The last JDDCP principle in the first group is *interoperability*, which requires that data citation methods are flexible enough to operate through different communities and citation practices.

The second group of JDDCP principles defines some of the requisites for data citation methods and systems; it is composed of *unique identification*, *access*, *persistence* and *specificity*. Unique identification indicates that a citation system must provide a methodology for identification that is machine-actionable and widely used within the community of reference. This principle is connected to: (i) *persistence*, which states that unique identifiers should persist even after the lifespan of data they are associated with; and (ii) *specificity*, which states that we should be able to identify specific data (even a data subset) to support a claim and that such data should be connected to provenance information useful for reconstructing their context. Persistence or *fixity* is a key factor for data citation because a data citation system should guarantee that a cited object is always available in the cited form across time. Ensuring the persistence of a database may not be enough (Thorisson, 2009), because to guarantee fixity of data citation we need database versioning systems; indeed, maintaining only the latest version of a dataset is not sufficient given that cited data may change or become unavailable across time without the knowledge of those who are citing it (Borgman, 2012b; Pröll & Rauber, 2013; Rauber, Ari, van Uytvanck & Pröll, 2016).

In a context where citation to big and evolving datasets has to be created and maintained over time, the requirements a data citation method should respect are becoming more demanding and complex than those defined by JDDCP. Unique identification is the most general and widely recognized requirement of a data citation method (Fenner *et al.*, 2016), but the use of persistent identifiers (PID) such as the Digital Object Identifier (DOI) has been highlighted as not being the "magic bullet" able to solve all the identification problems because we need to be able to uniquely identify subsets of data with *variable granularity* (Kafkas, Kim, Pi & McEntyre, 2015; Van de Sompel, 2012; Crosas *et al.*, 2015) since "different fields require different levels of detail to be able to reproduce data" (Mayernik, 2016); in other cases, we may need to identify sets of data aggregated from different sources. The granularity problem cannot be addressed by the use of data papers as proxies for the data to be cited (Candela, Castelli, Manghi & Tani, 2015) since they can only be used for citing a dataset as a whole; the same issue affects widely-used on-line resources, which associate a PID to a dataset and make it available and de-referenceable on-line, but do not provide variable granularity access and identification of subsets of the data.

This problem introduces both theoretical and practical (further discussed below) issues that exist also for textual citations. Indeed, in the context of traditional publications there are cases where it is customary to cite entire documents (e.g. scientific practice) or to cite single pages or passages (e.g. in the humanities). Nevertheless, in bibliometrics the convention is to aggregate documents by common elements (e.g. author or journal) and the unit of analysis is often the cited document (Borgman, 1990). On the contrary, for data citation there is not an agreed and shared protocol to be followed and being able to identify and cite a subset of data is a central problem that requires further research.

## How data citation: Data citation methods

From the analyses conducted so far, we can state that the ideal data citation system should uniquely identify a dataset and subsets of it with different levels of coarseness (*identification*), attribute the ownership and responsibility of the data with variable granularity to the right people/institutions (*attribution*), guarantee the persistence of the data being cited as well as the citations themselves (*fixity*), and automatically create complete and consistent citation snippets (*completeness and consistency*) according to community practices and shared *metadata standards*.

This citation system should be flexible enough to accommodate diverse requirements and citation practices across different disciplines as well as heterogeneous data models such as the relational model, the XML hierarchical model, or the RDF graph model.

From case to case the objects being cited within a given dataset may vary considerably; indeed, a data citation system needs to be able to cite:

- A *single resource* such as a complete dataset or a specific resource within a dataset; e.g. an RDF resource identified by a URI, an XML node identified by a single path from the tree root, a single tuple within a table in a relational database.

- A *subset/selection of resources* such as a selection of resources or portions of resources within a dataset; e.g. a bunch of RDF statements (e.g. a subgraph of an RDF dataset), an XML sub-tree or a set of tuples selected from a relational table.

- An *aggregation of resources* such as the union or join of resources coming from different parts of a dataset or even different datasets; e.g. an RDF graph created by joining together two RDF sub-graphs, a bunch of XML subtrees spanning one or more files, a set of relational tuples obtained by joining tables within the same database or from different databases.

In Table II, we report the main characteristics of the data citation systems and methods proposed in the literature according to the aspects outlined above. We highlight how these methods address the problem of identification, the automatic creation of citation snippets (i.e. automatic or manual), the coarseness of the produced citations (i.e. single resources, subsets or aggregations) and if these methods are general or defined for specific data models.

Table II Summary of the main characteristics of the state-of-the-art citation methods. We take into account the identification method, how the citation snippet is created, how fixity is handled, the level of coarseness of data citation and the data type. N/A indicates that the specific characteristic has not been taken into account by the specific solution.

| Reference | Identification method | Citation snippet creation method | Coarseness of the citation | Data model/type |
|---|---|---|---|---|
| (Alawini *et al.*, 2017) | PID (URI) | Automatic, view based | Single resource | RDF |
| (Bandrowsy *et al.*, 2015) | PID (RRID) | Manual/Template based | Dataset | Any data type |
| (Buneman & Silvello, 2010) | PID + path to node | Automatic, rule-based | Single resource | XML |
| (Buneman, Davidson & Frew, 2016) | PID + path to node | Automatic, view-based | Multiple resources | XML |

                                                    

| (Crosas, 2011) DataVerse | PID + UNF | Manual | Single resource | Any data type |
|---|---|---|---|---|
| (Davidson, Deutsch, Milo & Silvello, 2017) | N/A | Automatic, view-based | Multiple resources | Relational DB |
| (Groth *et al.*, 2010) Nanopublication | PID | Manual | Single statement (triple) | RDF |
| (Honor *et al.*, 2016) | PID | Manual/Landing page | Multiple resources | File system |
| (Rauber, Ari, van Uytvanck & Pröll, 2016) | Queries used as proxy for data | Manual/Metadata based | Multiple resources | Any data type. CSV and relational DB implementation. |
| (Silvello, 2015) | Named-graphs | Manual | Multiple resources | RDF |
| (Silvello, 2017) | PID + path to node | Automatic, machine learning-based | Single resource | XML |
| (Zwölf, Moreau & Dubernet, 2016) | Queries used as proxy for data | Manual/Landing page (DUI) | Multiple resources | Relational DB |

*Identification*

Most of the solutions proposed in the literature use PID for addressing the identification problem. As we highlighted above, the use of PID is a viable solution for citing a dataset as a whole, but it is not straightforward to use them for citing subsets or aggregations of data. In several cases, identification is provided to a high level of coarseness and identifying data with variable granularity is not possible. For instance, Bandrowsy *et al.* 2015) promotes the use of Research Resource Identifiers (RRID), which are unique persistent identifiers assigned to scientific resources. In particular, they have been used with biological resources such as antibodies, model organisms, and tools (software, databases, services). RRID are meant to identify resources at a high level of granularity and "*for software and databases, we elected to identify just the root entity and not a granular citation of a particular software version or database*".

Van de Sompel (2012) outlines two options to address the identification problem: to mint a new PID for each segment of data that need to be cited, or to mint a single PID for the dataset and to store the user query selecting or aggregating data.

The first solution has been chosen by DataVerse (Crosas, 2011,) which assigns a DOI to a dataset and generates a UNF (Universal Numerical Fingerprint) for each data segment to be cited within the dataset. A UNF is a PID assigned to specific digital objects or subsets within a dataset and it allows for identifying data with different granularities. This system needs to be extended/revised to handle the citations of big dynamic data in order to be able to cite a subset of data on (i) selected variables and observations for large quantitative data; (ii) time-stamp intervals; and, (iii) spatial dimensions (Crosas *et al.*, 2015) and also to be able to handle queries to the data. The use of DOI rose some concerns about the cost of minting identifiers for large data archives (Henderson & Kotz, 2015).

Buneman & Silvello (2010) and Silvello (2017) discussed the case of XML data, where every single node of an XML file can be a citable unit. In this case, a PID can be assigned to every XML file in a collection, but assigning a PID to every node would be unsustainable if not unfeasible. Buneman & Silvello (2010) proposed to identify a single node within an XML file by associating the PID identifying the XML file with the unique key identifying a specific node within the given XML file – e.g. the path of the node from the root of the file. Silvello (2015) discussed the case of RDF datasets, where we may need to cite a set of RDF statements (e.g. a subgraph of the whole RDF dataset at hand). In this case, every single node in a RDF graph is uniquely identified by a URI (i.e. a PID), but we cannot possibly assign a URI for every possible aggregation of nodes or statement in a dataset. The proposed solution is based on the use of named graphs (Klyne, Caroll & McBride, 2014), which are a set of connected RDF statements identified by a global PID – i.e. an URI, which is the name of the graph.

Groth *et al.* (2010) proposed the nanopublication model where research data that may need to be cited (or for which provenance and attribution information have to be held for any other motive) are modeled as single statements in a subject-property-object form, as in an RDF triple. The RDF triple is uniquely identified by a URI associated to the triple as in named graphs. The nanopublication model defines the minimum citable unit to be a triple; even though, it is not straightforwardly extensible to identify an aggregation of statements, the nanopublication model is widely used by the bioinformatics research community (Clark, Ciccarese & Goble, 2014; Mina *et al.*, 2015) and there are some studies about its adoption in the humanities (Gradmann, 2014; Golden & Shaw, 2016).

Pröll & Rauber (2013) and Rauber, Ari, van Uytvanck & Pröll (2016) proposed an identification method based on assigning PID to queries, which are used as proxies for the data subset to be cited. The access to data subset is allowed by re-issuing the stored query and a citation is associated with the PID of the query identifying the data. This method is flexible and in principle works for any data format and model; it requires the development of an additional infrastructure for storing the queries and assigning a timestamp to each query on the basis of the last update of the whole database. Moreover, it requires the development of some methods to guarantee query uniqueness – the proposal is to re-write queries to a normalized form and then to compute a checksum to detect identical queries – and stable sorting to ensure that the sorting of the records in the returned dataset is unambiguous and reproducible. This method has been developed in the context of the RDA (RDA, 2015) and it has been implemented by several scientific datasets as detailed in (RDA, 2016).

Parsons (2012) analyzed the problem of identifying subsets of a given dataset in order to make them citable, stating that identifying the query with a PID and citing the query, or assigning a UNF to a data subset to be cited, as in the DataVerse model, are not viable solutions in the context of Earth Science where there are no queries to be used and there is no agreed canonical version for the data. The solution adopted in this field is to use spatial and temporal information (i.e. a structural index as they call it) to identify a specific subset, since in Earth Science these data are always available and are often sufficient to discriminate between two subsets of a given dataset. Hence, Klump, Huber & Diepenbroek (2016) discussed the use of PIDs to identify and cite time series and evolving datasets as in the case of Earth Science resources where "*time or space are dimensions of the data and a subset may be defined by a range or bounding parameters*". The solution proposed is to enrich a PID with "fragment identifiers" in the form `<PID>@<fragment identifier>` where the "fragment identifier" can be the version of a resource or the time interval to be considered.

The Virtual Atomic and Molecular Data Centre (VAMDC) (Zwölf, Moreau & Dubernet, 2016) foresees a hybrid solution for citing resource aggregations generated by issuing SQL queries to a federated database putting together many heterogeneous data sources. For identification purposes VAMDC stores the user queries as proxies for the data, but instead of using the query PID for the citations, it creates a file containing the extracted data, information about the databases (e.g. version and contributors) and bibliographical information about the scientific papers related to the data (i.e. a form of provenance). This file is associated to a Digital Unique Identifier (DUI), which points to a landing page containing all the information in the file. The idea is to use the DUI associated to the file generated for the user query as a citation and to use the landing page to get the information usually provided by the citation snippet.

In the context of neuroimaging, (Honor *et al.*, 2016) addressed the problem of citing an aggregation of resources (i.e. images) created on the fly by the users. Here, there are three levels of identification: the image level where each resource is identified by an individual PID (i.e. DOI), the project level where pre-defined sets of images are assigned with a PID and the functional level, where aggregations of resources defined on the fly by users are identified by a newly minted PID. This solution requires to provide a module for PID deduplication when two different PIDs are assigned to the same functional aggregation of resources.

*Metadata schemas*

One goal in automating data citations is to structure them in a standardized format so that they can be formatted according to predefined styles and they can be included in bibliography management tools.

There are several metadata formats for data citations (Altman & King, 2007; Green, 2010; Starr & Gastl, 2011) and they all share a common subset of elements (Ball & Duke, 2015): `author, publication date, title, edition, version, URI, resource type, publisher, unique number fingerprint` (a form of hash of the data), a persistent `URL` and `location`. DataCite (DataCite, 2016), a widely-recognized metadata format proposal for citing data, expands this common set of fields by adding others such as `subject, contributor, format, size, description, language, rights` and `funding reference`. DataCite is the citation format adopted by the Thompson Reuters DCI. (Parsons, 2012) stated that there are some differences between the DataCite schema and the metadata fields required for citing Earth Science Information Partners (ESIP) data: `Authors, Release Date, Title, Version, Editors, Archive` and/or `Distributor, Locator, Date` and `Time accessed, Subset used`. The main difference regards the specification of the subset of data being used. Starr *et al.* (2015) reports a different minimum set of metadata fields to be used to cite a dataset: `Dataset Identifier, Title, Description, Creator, Publisher/Contact, PublicationDate/Year/ReleaseDate, Version, Creator Identifier`(s) (optional), `License` (optional).

The Repository Early Adopters Experts Group (an initiative of FORCE11 and the NIH BioCADDIE[6] program) (Fenner *et al.*, 2016) stated that the minimum set of required metadata is `Dataset Identifier, Title, Creator, Data Repository, Publication Date, Version` and `Type`. (Bravo *et al.*, 2015) proposed an even smaller group of mandatory fields for citing biological resources: `Identification` (ID and/or DOI), `Bioresource name, Acronym` (if available); (if applicable) `Organization; Number of accesses/Date of last access`. By contrast, in the field of biodiversity Chavan (2012) underlined the need to develop two different metadata schemas, one related to published dataset citations and another for query-based citations; he also proposed six alternative data citation styles.

It is evident that one size does not fit all when it comes to metadata formats for data citation. Metadata formats for "traditional" resources in the digital library domain such as the well-known Dublin Core adopted a dynamic solution that could be a viable possibility also in the data citation context. Indeed, the Dublin Core metadata standard is composed of 15 base elements – i.e. the Simple Dublin Core – that can be extended by using the so-called "application profiles" that adapt the standard to the requirements of specific application domains (although an application profile is not only a set of extra metadata fields, it also comprises policies and guidelines defined for a particular application, implementation or object type). In a similar vein, DataVerse (Crosas, 2011) proposed to use the (Altman and King, 2007) metadata schema based on a minimal set of elements which can be extended from case to case.

Metadata schemas should also be employed to enable the machine readability of data citations. The FORCE11 Data Citation Implementation Group in 2014 focused on this aspect by proposing to extend the NISO Z39.96-2012 JATS XML standard for exchanging journal article content in a machine-readable fashion (Mietchen, McEntyre, Beck & Maloney, 2015). This effort led to the definition of the JATS 1.1d2 schema comprising several tags specifically defined for data citation. (see http://jats4r.org/data-citations).

---

[6] https://biocaddie.org/

*Completeness and consistency*

Completeness and consistency of data citations can be achieved by providing tools able to automatically create citation snippets for the cited data. The creation of citation snippets is often demanded of users; for instance, describing the research identification initiative, Bandrowsky *et al.* (2015) explained that in a federated search context the data creators are asked to insert the correct citation for the data that will be then used as a template provided to the users. Users are asked to build the citation snippet manually using the given template as a guide. Similar methods are employed by many other relevant scientific databases such as the Eagle-i open RDF dataset (Torniai, Bourges-Waldegg & Hoffmann, 2015), a "resource discovery" tool built to facilitate translational science research. All the resources in Eagle-i are citable and, on the one hand, the system provides a "cite me" functionality, which returns a text description about how to cite a given resource and where to retrieve all the relevant data; on the other hand, the citation snippet has to be composed manually by the user. Reactome, a free, open-source, curated and peer reviewed pathway database for bioinformatics funded by the US NIH (Croft, 2013) uses the same manual method to instruct users about how to cite a specific resource. DataVerse assigns metadata to each dataset in a static way; citations to subsets are created starting from the citation of the main dataset and by adding some specific information such as UNF, but there is not a dynamic creation of the citation snippet. In the nanopublication model (Groth *et al.*, 2010) each citable statement is enriched with provenance and attribution annotations that can be used to manually create a citation snippet for the statement.

It has been highlighted that the manual creation of citation snippets is a barrier towards an effective and pervasive data citation practice as well as a source of inconsistencies and fragmentation in the citations (White, 1982; Thorisson, 2009). Indeed, especially for big, complex and evolving datasets, users may not have the necessary knowledge (both of the domain and of the technical aspects) to create complete and consistent snippets.

Indeed, one of the RDA recommendations for citation systems (Rauber, Ari, van Uytvanck & Pröll, 2016) regards "automated citation text generation", which is required to lower the barrier for citing and sharing the data. The solution sketched out in this context is similar to the one proposed by Bandrowsky *et al.* (2015); during the ingest phase of a dataset, data creators are required to provide descriptive metadata that will be used to create citation snippets.

Recently, the problem of automatically creating citation snippets has been defined as a computational problem that requires new ideas from computer science to be addressed (Buneman, Davidson & Frew, 2016). Existing techniques to automatically build citation snippets consider a single data model at a time and/or a narrow subset of queries.

In this context, the XML model has been investigated thoroughly, principally because it is widely used for data sharing and exchange. Buneman & Silvello (2010) proposed a rule-based system creating citations by using only the information present in the data themselves. Given an XML file, this system requires that the nodes corresponding to citable units be identified (in advance) and tagged with a rule that is then used to generate a citation; the form of the rule is $C \leftarrow P$ where $C$ provides a concrete syntax of a human or machine-readable citation and $P$ is a path augmented with some decorated variables. The purpose of $P$ is to bind the decorated variables in order to use them in $C$. Once the given XML file has been prepared to be cited (i.e. the rules are in place), the citation of a citable unit within this file is generated by a conjunction of the rules retrieved from the node corresponding to the citable unit up to the root of the XML file. Basically, the system gathers all the rules in the path from the citable unit to the root and each rule contains a specification of the elements to be comprised in the citation that has to be generated. This system allows the automatic generation of both human- and machine-readable citations of single XML nodes.

Buneman, Davidson & Frew (2016) build on and extend this system by defining a view-based citation method for hierarchical data. The idea is to define logical views over an XML dataset, where each view is associated to a citation rule, which if evaluated generates the required citation snippet according to a predefined style. This method allows us to create citations for general queries and thus for single resources as well as subsets and aggregations of resources. Basically, given a dataset *D*, a query *Q*, a set of views *V* defined over *D* and a set of citation rules (one for each view), the first step executed by the system is to rewrite *Q* by using the views in *V* and thus obtaining a valid rewriting *Q'* of *Q*. As an example, we may find that *Q'* uses two views, say *{V1, V2}*; the second step is to take the views used in *Q'* and to evaluate the rules associated to them. In our example, we would have two rules, one associated to *V1* and one to *V2*, where each rule generates a valid citation, say *C1* and *C2*. The third step is to employ a predefined function to combine *C1* and *C2* into a single citation *C*. Davidson, Deutsch, Milo & Silvello (2017) and Davidson, Buneman, Deutsch, Milo & Silvello (2017) further formalized and extended this work for it to work with general queries for relational databases; they also highlight the technical relationships between data citation and provenance. In the same vein by exploiting database views, Alawini, Chen, Davidson, Portilho da Silva & Silvello (2017) proposed a system for citing single RDF resources. They developed a working system tested for the Eagle-i RDF dataset; which creates automatic citation snippets for any RDF graph node, formats them in JSON and in a given human-readable format and maintains fixity thanks to an ad-hoc RDF versioned data store. These approaches are well-defined and could be implemented to work rather efficiently, but their principal drawback is that the rules as well as the views have to be defined by hand and they require the active involvement of experts (data creators and data curators) who know both the dataset and specific query languages and data models.

Silvello (2017) proposed the learning to cite framework, which builds on the approaches described above, but overcomes their main drawback by automating the creation of rules/views. The basic idea is to learn a citation model directly from a given data collection by using a sample set of citation snippets for training purposes and then exploit such a model to build citations on the fly for any citable unit within that collection. This system produces citations for single resources (i.e. single XML nodes) which are not formally exact, but as close as possible to what is considered a "correct citation". Silvello (2017)'s method needs to be revised in order to face the challenging computational problems emerging when working with more general queries and with relational and graph-based models.

*Fixity*

Guaranteeing the persistency of the data being cited as well as of the citations themselves is a core aspect of data citation. The main technique addressing the fixity problem has been put forward by the RDA Working Group on Data Citation (Rauber, Ari, van Uytvanck & Pröll, 2016), which proposed a versioning system for relational databases; this versioning system works in concert with the query store and guarantees that the cited data are always retrievable over time even though the dataset has been modified.

However, in order to address fixity for all the different kinds of scientific datasets, we need to address versioning for all the relevant data models adopted to manage, share and access scientific datasets; i.e. at least relational databases, XML and RDF datasets. Versioning systems must work with time queries because data referred by a citation needs to be retrieved at the time the citation was issued. There are some viable solutions for relational databases (Bohlen, Gamper, Jansen & Snodgrass, 2009) that may require improvements from the efficiency viewpoint and an effective solution for XML (Buneman, Khanna, Tajima & Tan, 2004) that needs to be extended to work with time queries, whereas RDF-based versioning requires a major methodological advancement to be usable with time queries (Geerts, Unger, Karvounarakis, Fundulaki & Christophides, 2016) in the data citation context. A first RDF versioning system for RDF datasets is proposed in (Alawini *et al.*, 2017) even though time queries are not explicitly handled.

*Citation infrastructures*

                                                       

Data infrastructures storing, managing, providing access and preserving datasets are essential to the development of data citation. Only a little more than five years ago, De Waard (2012) pointed out that: "Overall, commercial publishers are not interested in owning or charging for research data or running repositories. There might be exceptions, but in general this is the case." King (2011) also underlined the need to cooperate across scholarly fields: "[..] unless we are content to let data sharing work only within disciplinary silos we need to develop solutions that operate, or at least interoperate, across scholarly fields." Lately the data repository situation has been evolving considerably and there are several viable alternatives providing services to store, manage and access open research data and cite them both at the private and public level and across scholarly fields. Indeed, we see the emergence of numerous general-purpose data repositories, at scales ranging from institutional, to open globally-scoped repositories such as Dataverse, FigShare, Dryad, Mendeley Data, Zenodo, DataHub, DANS, and EUDat (Wilkinson *et al.*, 2016). These infrastructures mint persistent identifiers (DOI), allow for versioning (in some cases) and accept a wide range of file formats (Amorim, Castro, Rocha da Silva & Ribeiro, 2016). Moreover, in most cases, they provide a ready-to-use text snippet (created from manually inserted metadata) for the datasets and links to the papers using the datasets. There are several other public and private services which mint and assign DOI to datasets, but do not store them and thus handle versioning "in-house"; relevant examples are the *DataCite initiative*[7] and the *Dataverse network*[8].

All the above-mentioned data infrastructures worked in the direction of allowing the citation of data at fixed levels of coarseness – i.e. dataset, database, single file – and do not allow variable granularity data citations. From this, it follows that citation snippets are statically assigned to the datasets and that there are no mechanisms in place to automatically create citation snippets for data subsets defined on the fly by users.

It emerges from the analysis we have conducted that data citation is not a by-product of data storage and access and cannot be handled just by adopting persistent identifiers or by requiring users to provide describing metadata to be used to create the citation snippets. Data citation requires the implementation of complex and comprehensive solutions that will have quite an impact on current data infrastructures.

## Conclusions

Profound changes have been taking place in the way research is done and science progresses. Data has become fundamental for building and interpreting new scientific results and a major player in scientific advancement. Experimental and observational data and scientific models are now "born digital", and the progressive shift towards the *data-intensive science discovery* paradigm impacts all scientific disciplines. Nevertheless, scientific data are still not considered first-class players in the system of science.

In this article, we investigated the main motivations why data citation is central for the development of science and we analyzed several studies from different scientific fields to understand their motivations and check for common elements and viewpoints. We have concluded that data citation is required to give credit to data creators and data curators; it is a common belief that credit and attribution serve as an incentive to scientists for sharing more and better data leading to the reproducibility of scientific experiments and results. To this end, data citation is central also for the development of executable papers that allow data to be connected with the results presented in a scientific paper; a further extension of this view is the possibility of sustaining and validating scientific claims in a dynamic and automatic way. Data citation has a major impact also for easing the discovery of hidden data sources by providing new access points to them. Public and private institutions investing in dataset creation and curation see data citation as an important means for measuring the impact of data; this is important also for directing investments and research funding. In fact, the diffusion of pervasive and consistent data citation practices could lead to the definition of new impact measures and methodologies for assessing the scientific production of individuals and research institutions. This aspect has risen several concerned voices asking for a more profound comprehension of the norms regulating the citation of data and analyzing author citation behavior before using data citations for bibliometrics purposes.

---

[7] http://www.datacite.org/
[8] http://dataverse.org/

Despite its relevance and the attention dedicated to the topic, data citation poses a number of questions that have only been partially addressed by the information and computer science communities:

- (*Identification problem*) What data can be cited? How do we define a data citable unit? How do we identify a single resource, a subset of resources and an aggregation of resources?

- (*Completeness problem*) When data is extracted from a large, complex, evolving database, how do we create an appropriate and informative citation for it? How do we guarantee the consistency of citation texts?

- (*Fixity problem*) How do we guarantee that cited data will be accessible in their cited form?

We analysed these questions in detail and provided an overview of the existing approaches to address these issues. We have seen that significant effort has been made to pursue the identification of data, mostly relying on PIDs. Other approaches use queries as proxies for identifying data, but general agreement about what is the best solution for this issue has not yet been reached.

The completeness problem is even more open for discussion, since most solutions still rely on different forms of manual (or computer-aided) creation of citation snippets, thus often leading to incomplete and inconsistent citations. Automatic methods are required to improve consistency and ease the citation process; the existing automatic methods are defined for specific data models or query types and need more study and work to be generalized in order to be used for any kind of scientific datasets and general queries.

Guaranteeing the persistency of both the cited data and citations is a fundamental task any citation system has to accomplish; up until now, few solutions provided a fully-fledged solution able to deal with fixity, and more research on versioning systems and time queries is required to deal efficiently with this issue.

There are other issues connected to data citation to which not enough attention has been dedicated in the past and that could open up new research directions in the field. One of these is c*itation identity* or the ability to discriminate between two citations referring to different data or different versions of the same data and between two different citations referring to the same data. Another challenge is *citation containment* or the possibility to determine if a citation refers to a superset or a subset of the data cited by another citation. We have also seen that only initial efforts have been conducted in the direction of defining a *theory of data citing* and that the understanding of the motivations for citing data and why data is cited, is a challenge of fundamental importance for the definition of reliable data citation metrics and data citation indexes.

Furthermore, easy to use citation tools have to be developed both from the data creators/curators/administrators and final users' viewpoint. The former requires tools for specifying what data can be cited (e.g. which are the citable units) and how data have to be cited (e.g. for defining rules, views, training sets, etc.). The latter require tools for actually citing the data and using the produced citations in their work. To this end, we may need to develop tools able to create citation for data selected through a graphical user interface and not necessarily through well-formed and formal queries.